\documentclass[aps,prd,twocolumn,superscriptaddress,showpacs,floatfix,letterpaper]{revtex4}
\usepackage[dvips]{graphicx}
\usepackage{xspace}
\usepackage{latexsym}
\usepackage{amssymb}
\usepackage{times}
\usepackage{mathptm}

\begin{document}

\title{Further study of neutrino oscillation with two detectors
in Kamioka and Korea}

\date{\today}

\newcommand{\bu}{\affiliation{Department of Physics, Boston University, Boston, MA 02215, USA}}
\newcommand{\ipmu}{\affiliation{Institute for the Physics and Mathematics of the Universe (IPMU), University of Tokyo, Kashiwa, Chiba 277-8582, Japan}}
\newcommand{\ncen}{\affiliation{Research Center for Cosmic Neutrinos, Institute for Cosmic Ray Research (ICRR), University of Tokyo, Kashiwa, Chiba 277-8582, Japan}}
\newcommand{\geneve}{\affiliation{Section de Physique, Universit\'{e} de Gen\`{e}ve, 1205 Gen\`{e}ve, Switzerland}}

\author{F.~Dufour}\bu \geneve
\author{T.~Kajita}\ncen \ipmu
\author{E.~Kearns}\bu \ipmu
\author{K.~Okumura}\ncen

\begin{abstract}
  This paper updates and improves the study of electron neutrino
  appearance in the framework of two far detectors at different
  oscillation maxima, specifically, Tokai-To-Kamioka-to-Korea. We used
  a likelihood based on reconstructed quantities to distinguish
  charged current $\nu_e$ interactions from neutral current
  background. We studied the efficiency of the likelihood for a 20\%
  photo-coverage in comparison of a 40\% photo-coverage. We used a
  detailed neutrino event simulation to estimate the neutral current
  background. With these analysis tools we studied the sensitivity of
  the proposed experiment to CP violation and mass hierarchy as a
  function of the off-axis angle.

\end{abstract}


\maketitle

\section{Introduction}

The $\theta_{13}$ angle, the hierarchy of the largest mass splitting,
and the CP violating phase are remaining undetermined quantities in
the PMNS matrix formulation of neutrino masses and
mixing~\cite{Maki:1962mu,Pontecorvo:1967fh}. CP violation in the
lepton sector and its relation to the matter anti-matter asymmetry of
the universe is especially interesting. To determine the CP violating
phase $\delta$ and the neutrino mass hierarchy, a powerful approach is
to measure electron neutrino appearance at both the first and second
oscillation maximum in a long baseline neutrino beam. Two different
approaches have been considered in order to make this comparative
measurement. One approach is to have two detectors at two different
baselines of the same beam, one positioned for optimum response at the
first oscillation maximum and the other positioned for optimum
response at the second oscillation maximum. The optimum response is
achieved using the off-axis technique~\cite{oatechnique} to produce a
narrow energy band. This is the approach of the Tokai to Kamioka to
Korea approach~\cite{Ishitsuka:2005qi}, henceforth referred to by the
unofficial acronym T2KK. Another approach is to use an on-axis
wide-band beam, and measure electron neutrino appearance from both the
first and second maxima with a single
detector~\cite{Barger:2006vy}. This is the approach employed by the
BNL-FNAL working group~\cite{Barger:2007yw} as a model for a long
baseline neutrino oscillation experiment in the United States.

\begin{figure}[tbp]
\begin{center}
    \includegraphics[width=3.4in]{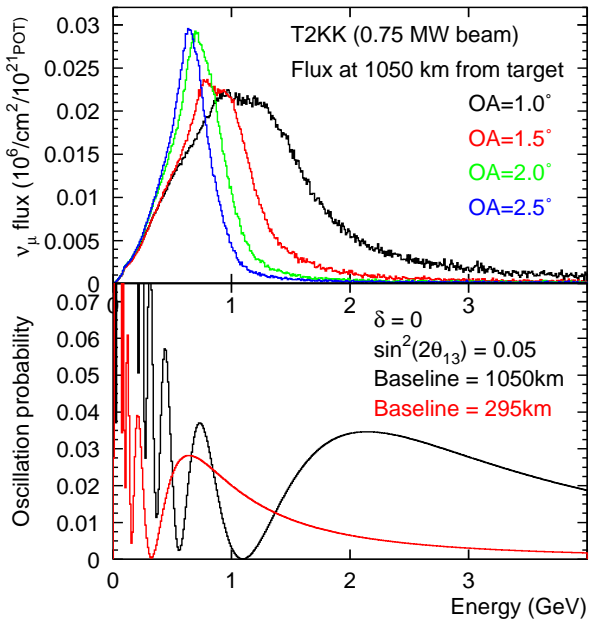}
\end{center}
  \vspace{-1.5pc}
  \caption{(Color online) Neutrino flux as a function of energy for
    several off-axis angle, and a 0.75MW beam at 1050km from the
    target.  For comparison, the $\nu_{\mu}$ $\rightarrow$ $\nu_e$
    probability, for two baselines considered for T2KK (295km and
    1050km). Neutrino mixing parameters are: normal hierarchy, $\Delta
    m^2_{(21,31)}=8.0\times 10^{-5},2.5\times 10^{-3} eV^2$, and
    $\sin^{2}2\theta_{(12,23)}=0.86,1.0$. We take the earth density to
    be constant and equal to 2.8 ${\rm g/cm^3}$.}
\label{fig:flux_t2kk}
\end{figure}

In the first published T2KK article~\cite{Ishitsuka:2005qi}, the
off-axis angle of the Korean detector was assumed to be fixed at
$2.5^{\circ}$. In this article, we study the sensitivity to CP
violation and mass hierarchy if we choose a smaller off-axis angle for
the location of the Korean detector, which blends the two approaches
described above. As one can see in Fig.~\ref{fig:flux_t2kk}, the
off-axis angle of $1.0^{\circ}$ results in a fairly wide band beam,
and we anticipate seeing electron neutrino appearance at both the
first and second maximum in the Korean detector. The detector at the
Kamioka location would remain at 2.5$^{\circ}$ off-axis, and be mainly
sensitive to the first oscillation maximum.

For this study, we assume an upgraded 1.66 MW J-PARC beam created from
40 GeV protons, running $10^7$ seconds per year. This is equivalent to
$2.59 \times 10^{21}$ protons-on-target (POT) per year. We assume 5
years of neutrino running and 5 years of anti-neutrino running.  The
$\nu_{\mu}$ flux observed at four different off-axis angles, at
1050~km from the target, is shown in Fig.~\ref{fig:flux_t2kk}. We also
assume two 0.27 Mton (fiducial volume) water Cherenkov detectors with
40\% photo-coverage. One of them would be located at Kamioka, at a
baseline of 295 km and at $2.5^{\circ}$ off-axis angle from the beam.
The second detector would be located in Korea at distances ranging
from 1000 to 1200 kilometers and off-axis angles ranging from
$1^{\circ}$ to $2.5^{\circ}$.  The beam intensity assumed for this
article is a factor of 2.4 lower than the beam assumed
previously~\cite{Ishitsuka:2005qi}. This more conservative beam power
is being considered for benchmark studies~\cite{NP08}. In addition we
assumed that a year of running is $10^7$ seconds instead of
$1.12\times 10^7$, so combining the change in beam intensity and
running time, the number of POT per year is a factor of 2.7 lower than
in previous studies.

Our tool for these studies is the fully reconstructed atmospheric
neutrino Monte Carlo sample from the Super-Kamiokande
experiment~\cite{Ashie:2005ik}, whereas the earlier paper used event
rates calculated for the T2K experiment scaled for distance. In order
to simulate the T2KK beam, we re-weight events in the Super-K
atmospheric neutrino Monte Carlo sample by the ratio of the T2KK flux
to the atmospheric flux.

\section{Signal VS. Background likelihood analysis}
\label{sec:likeli}

Our objective is to identify and reconstruct an excess of charged
current $\nu_e$ interactions in a nearly pure $\nu_\mu$ beam. We shall
be especially interested in quasi-elastic interactions such as $\nu_e
n \rightarrow e^- p$. In the experiment considered, the appearance
probability is a few percent at most, and only a small number of
events are anticipated above a non-negligible background. There are
three categories of background:

\begin{itemize}
\item $\nu_e$ beam: The irreducible background from electron neutrinos
in the beam flux regardless of neutrino oscillation. These come mainly
from muon decay and $K_{e3}$. 
\item Neutral current (NC): Background where the hadronic recoil of
neutral current interactions are misidentified as electron showers.
\item $\nu_{\mu}$ mis-ID: Background due to muons mis-identified as
  electron showers.
\end{itemize} 

\noindent The irreducible $\nu_e$ beam background is estimated from
the details of hadron production and muon decay in the beam Monte
Carlo.  We take as input the calculated flux from a beam simulation
assuming a graphite cylinder target, 30~mm in diameter and 900~mm in
length and a 130~m decay tunnel. These are the parameters of the T2K
experiment. The neutral current background mainly consists of hadronic
recoils with a single $\pi^0$. The $\pi^0$ decays into two photons and
if one of the photons is missed because of a very small energy or an
overlapping ring, then the event can be misidentified as a single
electromagnetic shower and therefore fake a $\nu_e$ CCQE event. The
dominant case is when one of the photons was missed because the energy
was too small. The $\nu_{\mu}$ mis-ID background consists of charge
current $\nu_{\mu}$ events where the Cherenkov ring from the outgoing
muon is mis-identified as an electron by the reconstruction
algorithm. This is the smallest source of background.

Since we are interested in $\nu_e$ appearance and especially $\nu_e$
undergoing quasi-elastic interactions, the events that we want to
select are fully contained inside the fiducial volume, have a single
Cherenkov ring identified as electron-like, and with no muon decay
electron present (a muon decay electron would signal a missed $\pi^+$
in a non-quasielastic final state).  These are referred to as
pre-cuts.  Before building the likelihood, we applied these pre-cuts,
in order to remove a significant part of the background.

The pre-cut efficiencies are listed in Table~\ref{tbl:bgpreeff}. The
NC efficiency is based on the total cross section for neutral current
interactions which includes a large component of neutrino-nucleon
elastic scattering. These are mostly unobserved in a water Cherenkov
detector. The NC background events that pass the pre-cuts are mostly
single-$\pi^0$ production.

\begin{table}[htbp]
\begin{center}
\begin{tabular}{|c||c|c|c||c|c|}
\hline
&\multicolumn{3}{c||}{Signal} & \multicolumn{2}{c|}{Background} \\
\hline
True $\nu$ energy & $\nu_e$ (avg) & QE $\nu_e$ & non-QE $\nu_e$ & NC 
& $\nu_{\mu}$ mis-ID\\
\hline
0 - 0.35 GeV &$95\%$ &$94\%$ & $53\%$ & $0.4\%$ & $0.5\%$ \\
0.35 - 0.85 GeV & $87\%$&$96\%$ &$49\%$ & $3\%$ & $0.4\%$\\
0.85 GeV - 1.5 GeV &$70\%$&$95\%$ &$43\%$& $8\% $ &$0.3\%$ \\
1.5 - 2.0 GeV &$58\%$&$91\%$ &$38\%$ &$11\% $ &$0.5\%$ \\
2.0 - 3.0 GeV &$51\%$ & $91\%$&$35\%$ & $11\% $ &$0.8\%$\\
3.0 - 4.0 GeV &$45\%$ &$90\%$ &$34\%$ & $12\% $ &$0.9\%$\\
4.0 - 5.0 GeV &$43\%$ &$90\%$ &$33\%$ & $13\% $ &$1.0\%$\\
5.0 - 10.0 GeV & $37\%$&$86\%$ &$29\%$ & $10\% $ &$1.4\%$\\
\hline
\end{tabular}
\caption {\label{tbl:bgpreeff} Efficiency of pre-cuts as applied
to neutrino interactions in the fiducial volume of the
Super-Kamiokande detector simulation. The charged current
$\nu_e$ interactions are broken down separately for quasi-elastic
and non-quasi-elastic samples. The NC sample includes elastic
scattering in the denominator of the efficiency calculation.}
\end{center}
\end{table} 

After applying pre-cuts, we make the final event selection using a
likelihood based on several event characteristics and using the ROOT
package TMVA~\cite{Hocker:2007zz}. This is a similar approach to one
previously studied by others~\cite{Yanagisawa:2007zz}. We reconstruct
the neutrino energy assuming quasi-elastic interactions. This depends
on particle masses, the reconstructed momentum and energy of the
outgoing lepton, and the angle between the outgoing lepton direction
and the known neutrino beam direction ($\theta_{\nu e}$):

\begin{equation}
E_{rec}=\frac{m_n E_e -m_e^2/2}{m_n - E_e +(P_e \cos{\theta_{\nu e}})}.  
  \label{eq:erec}
\end{equation}

\noindent The variables that are used in the likelihood can be divided
into three categories:
\begin{itemize}
\item Basic Super-Kamiokande event parameters:
  \begin {itemize}
    \item The ring-finding parameter used to count rings
    \item The $e$-like/$\mu$-like particle identification parameter
  \end {itemize}
\item Light-pattern parameters used for $\pi^0$ finding:
  \begin {itemize}
    \item The $\pi^0$ mass
    \item The $\pi^0$ likelihood
    \item The energy fraction of the $2^{nd}$ ring
  \end {itemize}
\item Beam related variable:
  \begin {itemize}
    \item The angle between the outgoing lepton and the beam direction
  \end {itemize}
\end {itemize}

\begin{figure}[htbp]
  \begin{center}
    \includegraphics[width=3.4in]{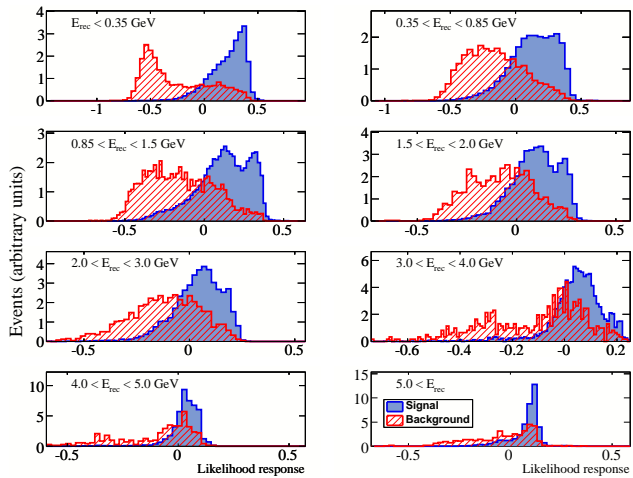}
  \end{center}
  \vspace{-1.5pc}
  \caption{(Color online) Combined likelihood distribution from 6
    input variables, shown separately for 8 energy bins. Charged
    current $\nu_e$ signal is shown in blue (filled), and the
    background is red (hatched). The events used have passed the
    defined pre-cuts.}
  \label{fig:like}
\end{figure}

\noindent We already cut on the ring parameter and the PID parameter
in the set of pre-cuts (Table~\ref{tbl:bgpreeff}). Here we used the
continuous distribution of these parameters as input to the
likelihood. There are three variables related to a specialized fitter
({\tt POLfit} for Pattern-Of-Light fitter) used to select single
$\pi^0$ events~\cite{Barszczak:2005sf}. The output of this fitter
includes an overall likelihood as well as the best fit mass and energy
fraction of the two gammas from $\pi^0$ decay.  We also use one
variable that requires knowledge of the beam direction, and therefore
is not a standard SK variable for atmospheric neutrino analysis. For
that variable, we had to use the MC truth information about the
neutrino direction in the simulated atmospheric neutrino Monte Carlo
sample.  Unlike the accelerator-based experiment, these events are
simulated over a wide-range of incident angles. However, the Super-K
detector has uniform response. The distributions of the combined
likelihood for each energy bin is shown in Fig~\ref{fig:like}. The
separation between signal and background is striking at low energies
but becomes worse at higher energies.

To choose where to cut on the likelihood variable, we compute the
signal over square root of background, $S/\sqrt{B}$ for several
positions of the cut. We tested cuts that range from keeping 10\% of
signal to keeping 100\% of the signal (at the expense of increasing
background). We also varied the off-axis angle and considered separate
energy bins. We found that keeping a large fraction of signal, 80\%,
maximizes $S/\sqrt{B}$. The energy dependent efficiencies for an 80\%
likelihood cut is given in Table~\ref{tbl:bglikeeff}.

\begin{table}[htbp]
\begin{center}

\begin{tabular}{|c||c|c|c|}
\hline 
& \multicolumn{3}{l|}{Cut that keeps 80\% of signal} \\ 
\hline
Energy (rec) & ~~~~$\nu_e$~~~~& ~~~~NC~~~~ & $\nu_{\mu}$ mis-ID\\ 
\hline 0 - 350 MeV &$80\%$ &$15\%$ & $15\%$ \\ 
350 - 850 MeV & $80\%$ &$25\%$ & $40\%$\\ 
850 MeV - 1.5 GeV &$80\%$ &$28\%$ &$30\%$ \\ 
1.5 - 2.0 GeV &$80\%$ &$30\%$ &$32\%$  \\ 
2.0 - 3.0 GeV & $80\%$ &$40\%$& $18\%$\\ 
3.0 - 4.0 GeV & $80\%$&$50\%$ & $28\%$\\ 
4.0 - 5.0 GeV & $80\%$&$65\%$ & $55\%$\\ 
5.0 - 10.0 GeV &$80\%$ &$45\%$ & $18\%$ \\ 
\hline 
\end{tabular}
\caption {\label{tbl:bglikeeff} Efficiency for the likelihood cut that
  keeps 80\% of the signal. These efficiencies are calculated for
  events which have already passed the pre-cuts, and are calculated
  based on reconstructed energy.}
\end {center}
\end {table}

\subsection{Photo coverage}
\label{sec:coverage}

Due to the accident that happened in November 2001, where about half
of the Super-Kamiokande phototubes were destroyed, Super-K has run
with both 40\% and 20\% photo-coverage, and has atmospheric neutrino
Monte Carlo samples for both conditions. It was therefore easy to
repeat our studies for a detector with a 20\% photo-coverage.  Overall
we found that both the pre-cuts and the likelihood are nearly as
efficient in a detector with 20\% coverage as they are in a detector
with 40\% coverage. The comparison of the pre-cuts is presented in
Table~\ref{tbl:covpreeff} and the comparison of the likelihood is
presented in Table~\ref{tbl:covlikeeff}. 

\begin{table}[tbp]
\begin{center}
\begin{tabular}{|c||c|c||c|c||c|c|}
\hline
&\multicolumn{2}{c||}{$\nu_e$} & \multicolumn{2}{c||}{NC} & \multicolumn{2}{c|}{$\nu_{\mu}$} \\
\hline
&\multicolumn{2}{c||}{Photo-cov.} & \multicolumn{2}{c||}{Photo-cov.} & \multicolumn{2}{c|}{Photo-cov.} \\

Energy (true) & 40\% &  20\% & 40\% & 20\% &  40\% &  20\%\\
\hline
0 - 350 MeV &$95\%$&$93\%$  &$0.4\%$ &$0.3\%$&$0.5\%$ &$0.5\%$ \\
350 - 850 MeV & $87\%$ & $86\%$ & $3\%$ & $3\%$ &$0.4\%$ &$0.5\%$\\
850 MeV - 1.5 GeV &$70\%$&$70\%$ &$8\%$ &  $9\%$&$0.3\%$ &$0.5\%$\\
1.5 - 2.0 GeV &$58\%$&$57\%$ &$11\%$ &$12\%$ &$0.5\%$ &$0.5\%$\\
2.0 - 3.0 GeV & $51\%$& $49\%$ &$11\%$ &$12\%$ &$0.8\%$ &$0.6\%$\\
3.0 - 4.0 GeV & $45\%$ & $45\%$&$12\%$ & $12\%$&$0.9\%$ &$0.9\%$\\
4.0 - 5.0 GeV & $43\%$ & $45\%$ &$13\%$ &$13\%$ &$1.0\%$ &$0.9\%$\\
5.0 - 10.0 GeV & $37\%$& $37\%$ &$10\%$ & $12\%$&$1.4\%$ &$1.0\%$\\
\hline
\end{tabular}
\caption {\label{tbl:covpreeff} Pre-cut efficiency for two
  photo-coverage: 40\% (SK-I) and 20\% (SK-II). The NC sample includes
  elastic scattering in the denominator of the efficiency
  calculation.}
\end{center}
\end{table}

\begin{table}[htbp]
\begin{center}
\begin{tabular}{|c||c|c||c|c|}
\hline
&\multicolumn{2}{c||}{$\nu_e$} & \multicolumn{2}{c|}{NC}  \\
\hline
&\multicolumn{2}{c||}{Photo-coverage} & \multicolumn{2}{c|}{Photo-coverage}  \\
Energy (rec) & ~~$40\% $~~ & $20\%$ & ~~$40\% $ ~~& $20\% $\\
\hline
0 - 350 MeV &$80\%$&$80\%$  &$15\%$ &$15\%$ \\
350 - 850 MeV & $80\%$ & $80\%$ & $25\%$ & $24\%$ \\
850 MeV - 1.5 GeV &$80\%$&$80\%$ &$28\%$ &  $25\%$\\
1.5 - 2.0 GeV &$80\%$&$80\%$ &$30\%$ &$35\%$ \\
2.0 - 3.0 GeV & $80\%$& $80\%$ &$40\%$ &$40\%$ \\
3.0 - 4.0 GeV & $80\%$ & $80\%$&$50\%$ & $42\%$\\
4.0 - 5.0 GeV & $80\%$ & $80\%$ &$65\%$ &$50\%$ \\
5.0 - 10.0 GeV & $80\%$& $80\%$ &$45\%$ & $45\%$\\
\hline
\end{tabular}
\caption {\label{tbl:covlikeeff} Likelihood efficiency for two
  photo-coverages: 40\% (SK-I) and 20\% (SK-II). The likelihood
  cut keeps 80\% of signal.}
\end{center}
\end{table}

\section{How to Compute the Background Spectrum}
\label{sec:bg}

As mentioned in Section~\ref{sec:likeli}, there are three categories
of background: $\nu_e$ beam background ($\nu_e$ beam), neutral current
background (NC), and charged current $\nu_{\mu}$ mis-identified
background ($\nu_{\mu}$ mis-ID). To simulate the background in the
long baseline beam experiment, we used the SK atmospheric Monte Carlo
as follows:

\begin{itemize}
\item We ran over the atmospheric SK Monte Carlo, and kept events
  which passed all the pre-cuts.

\item We applied the likelihood efficiency corresponding to the right
  background type ($\nu_e$, $\nu_{\mu}$ mis-ID or NC) and using the
  reconstructed energy.  This takes care of the likelihood efficiency,
  and also the energy resolution of the detector since we use
  reconstructed energy.

\item We re-weighted this background spectrum by the ratio of the beam
  $\nu_{\mu}$ flux to the atmospheric flux.
 
\item We normalized the final background spectrum in order to account
  for the running conditions of the experiment: volume of detector,
  beam power, etc.
\end{itemize}

\begin{figure}[tbp]
\includegraphics[width=2.8in]{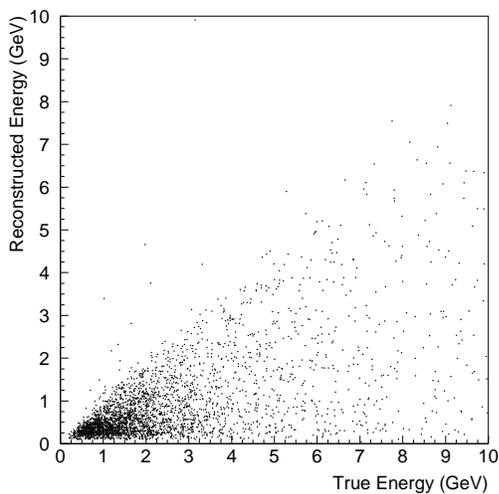}
\caption{Smearing matrix for neutral current events: the result of
  energy reconstructed using Eqn.~\ref{eq:erec} versus true neutrino
  energy.}
\label{fig:ncsmear}
\end{figure}

\noindent It is important to consider the neutral current background
properly since its energy response is very uncorrelated, as can be
seen in Fig.~\ref{fig:ncsmear}.

\section{Off-axis angle analysis}
\label{sec:off_axis}

\begin{figure*}[htbp]
\begin{center}
{\hbox{\hspace{0.0in}
    \includegraphics[width=3.5in]{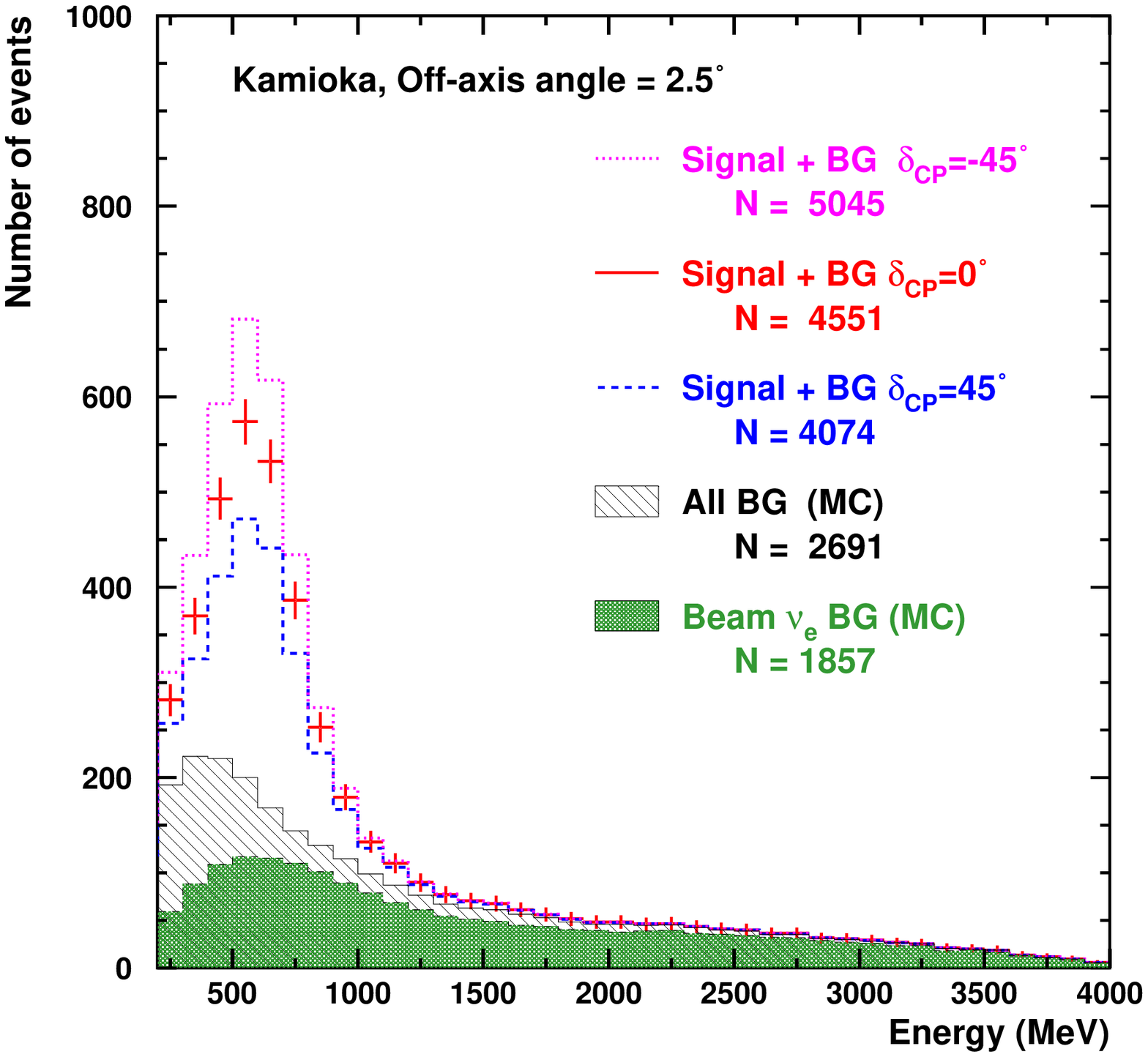}
}}
{\hbox{\hspace{0.0in}
    \includegraphics[width=3.5in]{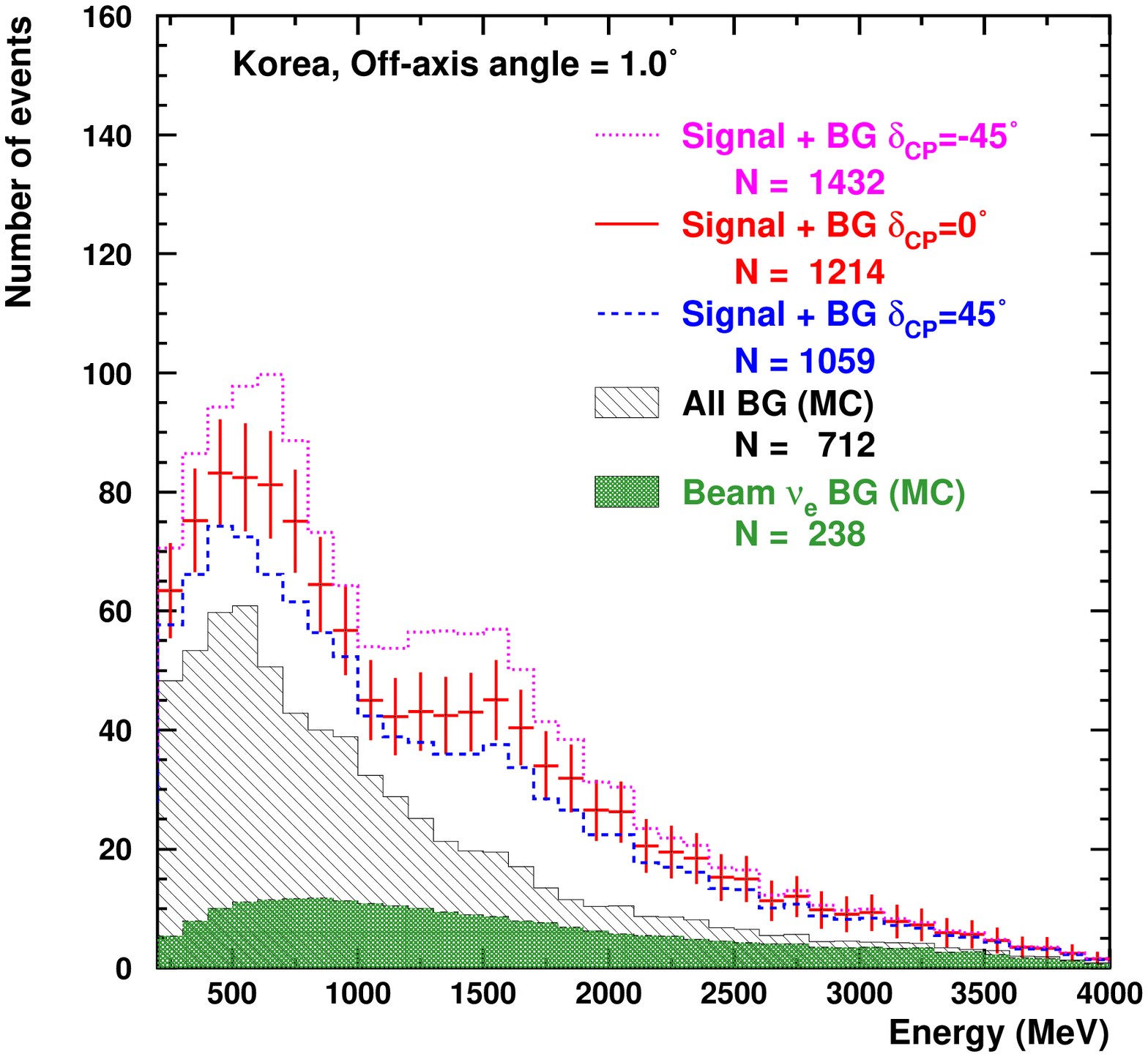}
\hspace{0.0in} \includegraphics[width=3.5in]{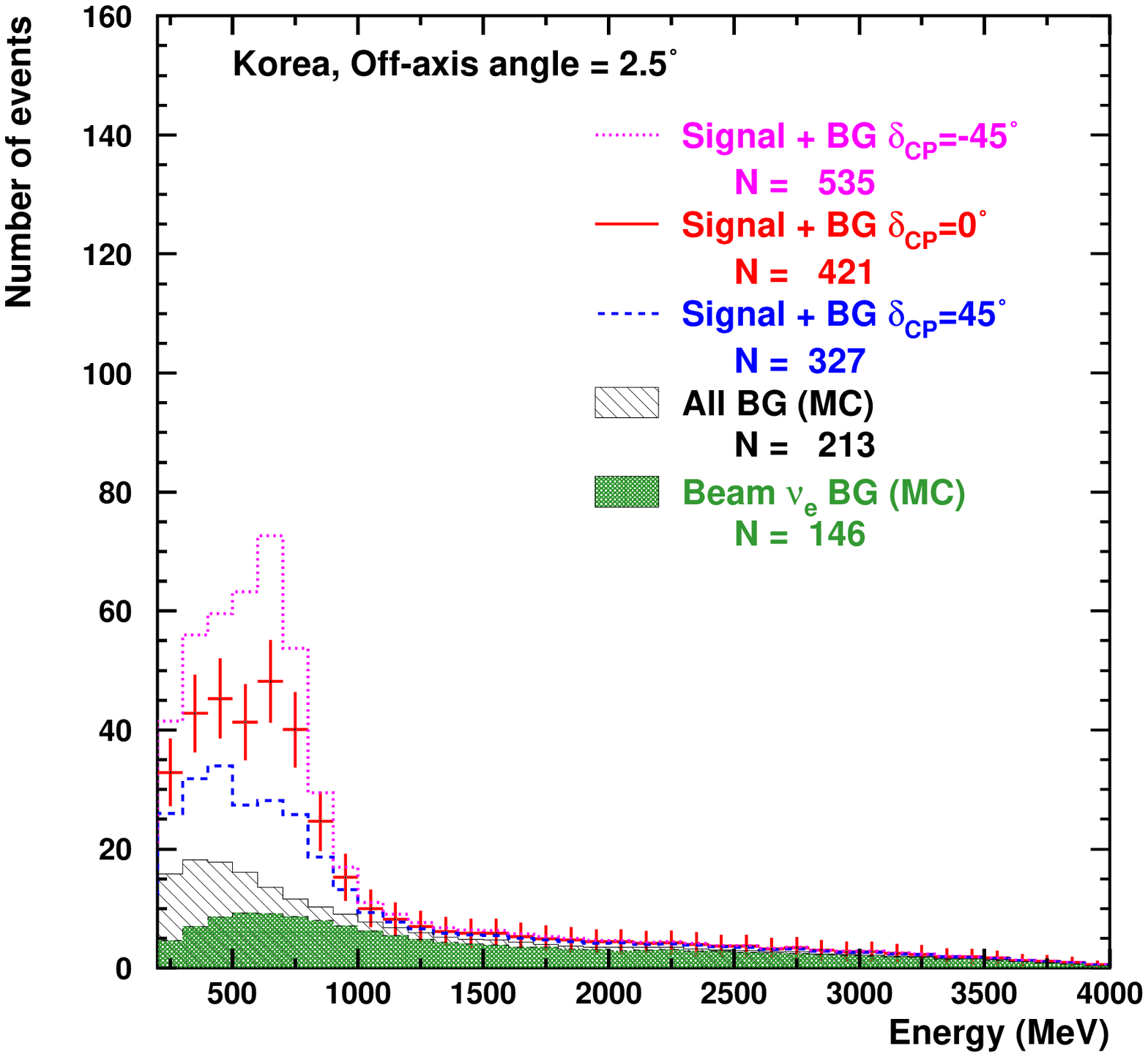}
}}
\end{center}
  \vspace{-1.5pc}
  \caption{(Color online) Reconstructed energy spectra at Kamioka
    (top), Korea $1.0^\circ$ off-axis (bottom left) and Korea
    $2.5^\circ$ off-axis (bottom right) for $\sin^2(2\theta_{13})$
    =0.04 and normal hierarchy. The remaining oscillation parameters
    are: $\Delta m^2_{(21,31)}=8.0\times 10^{-5},2.5\times 10^{-3}
    eV^2$ and $\sin^{2}2\theta_{(12,23)}=0.86,1.0$. Each plot is
    normalized to 5 years of running with neutrino, a 1.66 MW beam
    with 40 GeV protons and in a 0.27 Mton (FV) detector (i.e. $5
    \times 2.59 \times 10^{21}$ POT).}
\label{fig:spectrum_like80}
\end{figure*}

Using the cut on the likelihood that keeps 80\% of the signal, we
present in Fig.~\ref{fig:spectrum_like80} spectra at the Kamioka
location and at the Korean location for $1^\circ$ off-axis angle and
$2.5^\circ$ off-axis angle. We also present the sensitivity to mass
hierarchy and CP violation, for four different values of the off-axis
angle position of the Korean detector. The $\chi^2$ analysis used to
compute the sensitivity is similar to that previously
used~\cite{Ishitsuka:2005qi} and is defined as:

\begin{equation}
\label{eq:chisquare}
{\chi^2 = \sum_{k=1}^{N_{exp}} \left( \sum_{i=1}^{N_{Ebin}}
\frac{(N(e)^{obs}_i -N(e)^{exp}_i)^2}{\sigma^2_i}\right)
+ \sum_{j=1}^{15} \left( \frac{\epsilon_j}{\tilde{\sigma_j}}
\right)^2, }
\end{equation}

\noindent where

\begin{eqnarray}
N(e)^{exp}_i & = & N^{BG}_i \cdot (1+\sum_{j=1}^7 f^i_j \cdot \epsilon_j)
+ N^{signal}_i \cdot (1+ \sum_{j=8}^{13}f^i_j \cdot \epsilon_j) \nonumber \\
& + & N^{\Delta E~scale}_i \cdot (1+ \sum_{j=14}^{15} f^i_j \cdot \epsilon_j).
\label{eq:nexp}
\end{eqnarray}

\noindent Here, $N_{exp}$ is the number of ``experiments''. For
example if we have two detectors (Kamioka and Korea) and run with only
neutrinos then $N_{exp}=2$.  If we have two detectors but run with
neutrinos and anti-neutrinos then $N_{exp}=4$. Compared to the
publication of Ishitsuka {\it et al.}, we added two energy bins and
use events up to 3 GeV, which is relevant when the Korean detector is
located at small off-axis angles. So for this analysis, we have
$N_{exp}=4$ since we ran for neutrinos and anti-neutrinos and have two
detectors. We have 7 energy bins ($N_{Ebin}$): 400-500 MeV, 500-600
MeV, 600-700 MeV, 700-800 MeV, 800-1200 MeV, 1200-2000 MeV, 2000-3000
MeV. The sum over $j$ in Eq.~\ref{eq:chisquare} is the sum over the
systematic errors. We consider fifteen systematic errors
$\tilde{\sigma_j}$ in this study. They are presented in
Table.~\ref{tbl:sys_description} and they are split into three
groups. The first seven errors are on the background, the next six on
the signal, and the last two on the energy scale. The largest
systematic uncertainty comes from the signal normalization above $1.2$
GeV, and this is due to the uncertainty on the number of rings for
Multi-GeV electron-like events~\cite{Takenaga}.  The systematic
uncertainties were estimated using work by the Super-Kamiokande
collaboration~\cite{Ashie:2005ik,Takenaga,Maxim}.

\begin{table}[htbp]
\begin{center}
\begin{tabular}{|c|l|c|}
  \hline  Index&Systematic uncertainty & Value\\
  \hline
 1& BG normalization below 1.2 GeV (for Kamioka) & 5\% \\
 2& BG normalization above 1.2 GeV (for Kamioka) & 5\% \\
 3& BG normalization below 1.2 GeV (for Korea) & 5\% \\
 4& BG normalization above 1.2 GeV (for Korea) & 5\% \\
 5& BG normalization between $\nu_e$ and $\bar{\nu_e}$ below 1.2 GeV& 5\% \\
 6& BG normalization between $\nu_e$ and $\bar{\nu_e}$ above 1.2 GeV& 5\% \\
 7& BG spectrum (common for Kamioka and Korea) & 5\% \\
\hline
 8& Signal normalization below 1.2 GeV $\sigma(\nu_{\mu})/\sigma(\nu_e)$ & 5\% \\
 9& Signal normalization above 1.2 GeV $\sigma(\nu_{\mu})/\sigma(\nu_e)$ & 20\% \\
 10& $[\sigma(\nu_{\mu})/\sigma(\nu_e)]/[\sigma(\bar{\nu_{\mu}})/\sigma(\bar{\nu_e})]$ below 1.2 GeV
  &5\% \\
 11& $[\sigma(\nu_{\mu})/\sigma(\nu_e)]/[\sigma(\bar{\nu_{\mu}})/\sigma(\bar{\nu_e})]$ above 1.2 GeV
  &5\% \\
 12&Efficiency difference between Kamioka and & \\
 & Korea detector below 1.2 GeV & 1\%  \\
 13&Efficiency difference between Kamioka and & \\
 & Korea detector above 1.2 GeV & 1\%  \\
\hline
 14& Energy scale difference between Kamioka and & \\
& Korea detector  & 1\%    \\
 15& Energy scale difference between near and & \\
& (Kamioka/Korea) detector  &1\% \\
\hline
\end{tabular}
\caption {\label{tbl:sys_description} List of systematic uncertainties
  and their assumed values.}
\end{center}
\end{table}

\begin{figure*}[htbp]
\begin{center}
{\hbox{\hspace{0.0in}
    \includegraphics[width=3.5in]{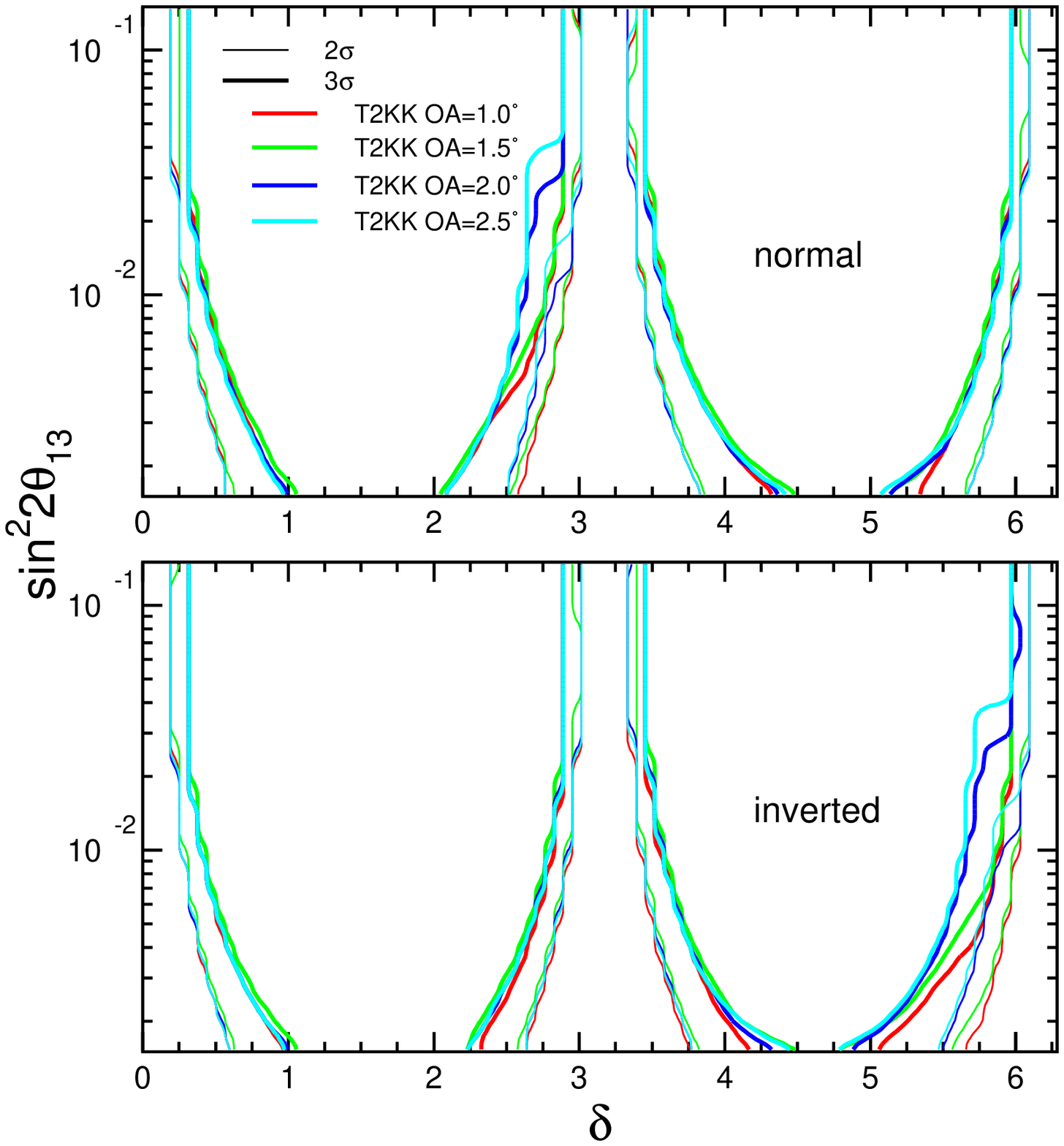}
\hspace{0.0in} \includegraphics[width=3.5in]{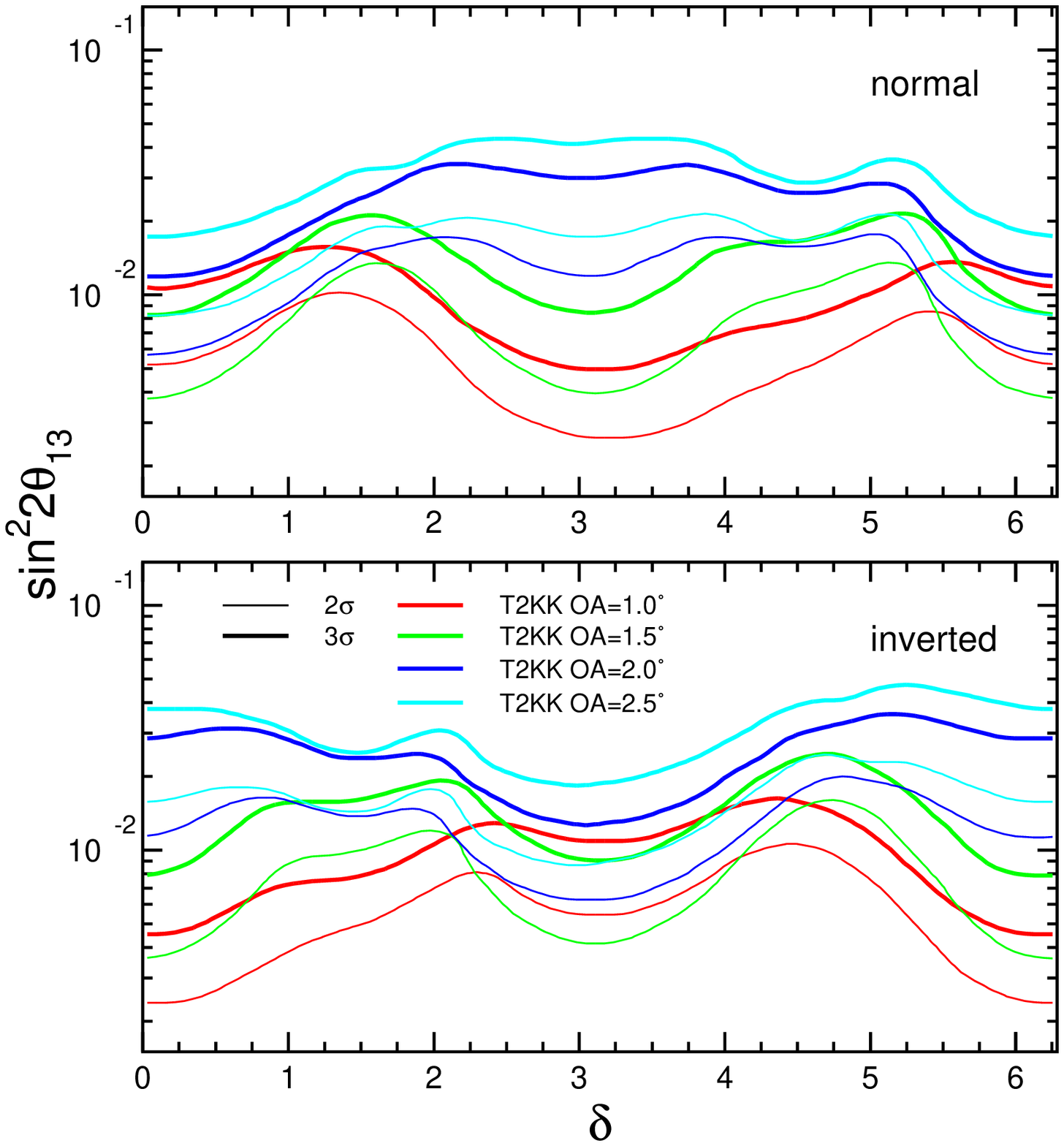}
}}
\end{center}
  \vspace{-1.5pc}
  \caption{Sensitivity to CP violation (left) and mass hierarchy
    (right) for different values of the off-axis angle. Other mixing
    parameters are the same as used in previous figures.  Each plot
    considers 5 years of running with neutrinos and 5 years with
    anti-neutrinos, a 1.66 MW beam with 40 GeV protons and in a 0.27
    Mton (FV) detector, i.e. $10 \times 2.59 \times 10^{21}$ POT. The
    parameter region above the curve is the region where we can
    determine CP violation or mass hierarchy to either 2 or 3
    $\sigma$.}
\label{fig:region-like80}
\end{figure*}

\begin{figure*}[htbp]
\begin{center}
{\hbox{\hspace{0.0in}
    \includegraphics[width=3.5in]{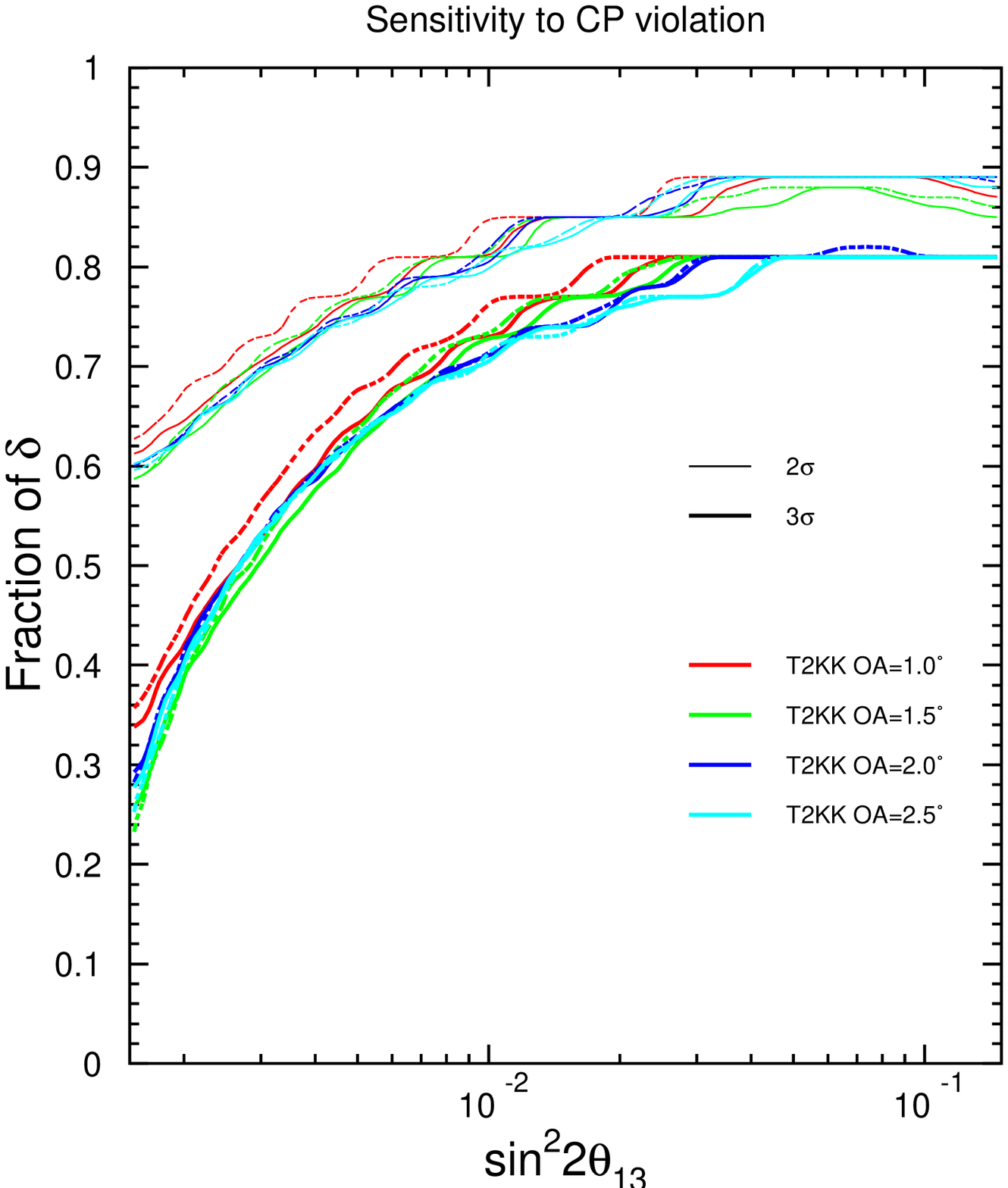}
\hspace{0.0in} \includegraphics[width=3.5in]{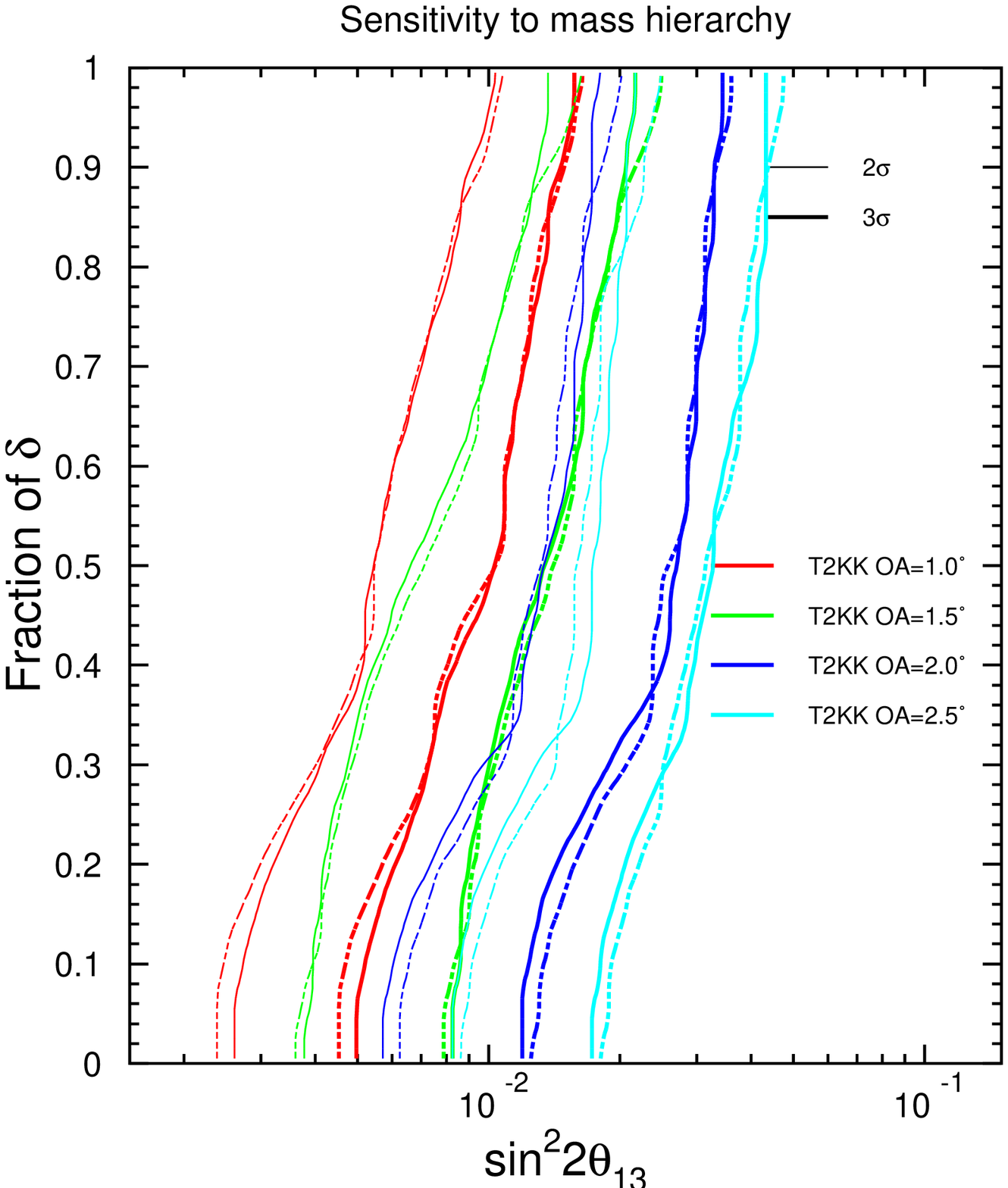}
}}
\end{center}
  \vspace{-1.5pc}
  \caption{Sensitivity to CP violation (left) and mass hierarchy
    (right) for different values of the off-axis angle. Other mixing
    parameters and the beam exposure are the same as used in previous
    figures. These plots show for what fraction of possible values of
    the CP phase $\delta$ we will be able to determine CP violation or
    mass hierarchy. The plain lines are for normal hierarchy while the
    dotted lines are for inverted hierarchy. If $\theta_{13}$ is large
    enough the mass hierarchy can always be determined whatever the
    value of $\delta$; this is not the case for establishing CP
    violation when $\delta$ approaches 0, $\pi$, or $2\pi$.}
\label{fig:fraction-like80}
\end{figure*}

The results for the mass hierarchy and CP violation sensitivity are
presented in Fig.~\ref{fig:region-like80} and
Fig.~\ref{fig:fraction-like80}. We find that the best sensitivity to
both CP violation and mass hierarchy is achieved with the Korean
detector located at $1^\circ$ off-axis. The improvement in sensitivity
to CP violation is rather minimal, however the sensitivity to mass
hierarchy is improved by a factor of three compared to the original
configuration with the Korean detector located at $2.5^\circ$ off
axis. This is due to the information gained by including the first
oscillation maximum, with higher energy neutrinos, in the Korean far
detector.

We note that several improvements have been made since the T2KK
article published in 2005~\cite{Ishitsuka:2005qi}. Several minor
problems were fixed and the cut on the likelihood variable was
added. This allowed us to gain a significant number of signal
events. For example in the 350-850 MeV bin, the combined efficiency
(pre-cuts and likelihood) is 68\%, where in the same bin of
Ref.~\cite{Ishitsuka:2005qi} it was 40\%. In addition, the likelihood
cut allows us to increase $S/\sqrt B$. Again for the 350-850 MeV bin,
the $S/\sqrt B$ was increased by about 20\%. If we had run with the
same number of protons-on-target as the authors of the 2005 paper, the
sensitivity would be a factor of two better than what we are reporting
for $2.5^\circ$ off-axis angle. Conversely, with the conservative
benchmark beam power of 1.66 MW instead of 4MW, our sensitivity with
the Korean detector located at $2.5^\circ$ off-axis is roughly
equivalent to that of the 2005 paper.

\section{Conclusions}

We have presented an updated and improved study of long baseline
neutrino oscillation with a detector in Kamioka and a second in Korea.
Using pre-cuts and a likelihood designed to reject neutral current
background while keeping charged current quasi-elastic events, we were
able to increase the amount of signal that we keep in the main signal
bin (350-850 MeV) from about 40\% to 68\%, and we were able to remove
more background than what was done before. We found that the
effectiveness of the cuts and likelihood was relatively undiminished
when applied to a detector with 20\% rather than 40\%
photo-coverage. We found that the best location among the possibilities
we explored for the Korean detector is $1.0^{\circ}$, which is more
on-axis than previously considered, and allows a somewhat wider band
neutrino energy spectrum. This improved the T2KK sensitivity by a
factor of two compared to what was published previously, even after
taking a more conservative number of POT per year to be a factor of
three lower. With an experiment configured at $1.0^{\circ}$, and a
benchmark beam power of 1.66 MW, the neutrino mass hierarchy should be
revealed if $\sin^2 2\theta_{13}$ is larger than $10^{-2}$ for a wide
range of $\delta$. CP violation would be detected at 3$\sigma$
significance for 70\% of possible values of $\delta$.

The authors gratefully acknowledge the work of the Super-Kamiokande
collaboration in developing the tools used for this analysis, however
the results and their interpretation are the responsibility of the
authors of this paper. We are thankful to the authors of
Ref.~\cite{Ishitsuka:2005qi} who generously provided their software.
We are grateful for financial support by the United States Department
of Energy and the Grant-in-Aid Scientific Research by the Japan
Society for the Promotion of Science.

\bibliography{t2kk08_paper}
\end{document}